Article

# Compact orthogonal view OCT: clinical exploration of human trabecular meshwork and cornea at cell resolution

Viacheslav Mazlin[1,*], Kristina Irsch[2,3], Vincent Borderie[3], Christophe Baudouin[3], Michel Paques[3], Mathias Fink[1], Kate Grieve[2,3] and A. Claude Boccara[1]

[1]*Institut Langevin, ESPCI Paris, PSL University, CNRS, 1 Rue Jussieu, Paris, 75005, France*

[2]*Vision Institute, CIC 1423, Sorbonne University, UMR_S 968 / INSERM, U968 / CNRS, UMR_7210, 17 Rue Moreau, Paris, 75012, France*

[3]*Quinze-Vingts National Eye Hospital, CIC 1423, 28 Rue de Charenton, Paris, 75012, France*

**Corresponding author (marked with*):**

VM: mazlin.slava@gmail.com

## Abstract

Glaucoma remains among the leading causes of blindness despite many treatment options available today. Effective treatment requires early diagnosis, which is difficult to achieve with existing imaging technologies that detect the already inflicted irreversible posterior eye damage. One improved approach could be to detect the changes in trabecular meshwork of the anterior eye that precede the intraocular pressure buildup and vision loss. Here we report a device that enables in vivo exploration of trabecular meshwork at cell resolution. The instrument combines two orthogonal views: en face view of time-domain full-field OCT for high-resolution and cross-sectional view of spectral-domain OCT for aligning to the meshwork. The device resolves micron-level corneoscleral and uveal pores, meshwork beams and internal elastic fibers, ciliary body tendons and trabecular cell nuclei. Capable of detecting blockage of trabecular pores, orthogonal view OCT presents a promising solution to the long-standing problem of early glaucoma diagnosis. Future clinical adoption of the device is facilitated by the compact footprint and broad functionality that includes corneal imaging in healthy subjects and patients (with keratoconus, Fuchs' endothelial dystrophy) using the same design.

# Introduction

Glaucoma is a leading cause of irreversible blindness affecting 60 million people worldwide [1]. Glaucoma is not a single disease but a group of diseases, all of which lead to a progressive loss of retinal ganglion cells that play essential role in visual function. The cell loss has been attributed to the diverse factors including impaired blood supply (ischemia), inflammatory damage, withdrawal of cell protection (neurotrophic) mechanisms and excessive production of neurotoxic molecules (excitotoxicity) [2], however mechanical damage through high intraocular pressure (IOP) [3] remains the most common one present in about 80% of glaucoma cases [4]. Although numerous medication therapies exist for lowering IOP [2], the decision to initiate them should be taken only following the definite diagnosis of glaucoma, as the therapies have far-reaching consequences for the patient: altered quality of life from the life-long medication use, side effects and significant costs. Unfortunately, IOP measurements alone have low specificity in discriminating glaucomatous eyes from normal. Several retinal imaging and functional instruments (for example, fundus camera, automated perimetry) can be used in junction with IOP measurements to confirm glaucoma diagnosis. However, they detect the consequence of the prolonged high IOP - the altered retinal and optic nerve structures. The better approach would be to find the precise cause of increased IOP in the first place, evaluate the risks of glaucoma associated with this cause and select a suitable treatment before any posterior eye damage happens.

High IOP is almost always due to poor outflow of aqueous humour through the trabecular meshwork, located at the angle of the anterior chamber [5,6]. Clinical conditions, where trabecular meshwork is blocked by the iris (angle-closure glaucoma), can be directly viewed with gonioscopy. On the other hand diagnosis of a more common open-angle glaucoma (OAG) (75% of all glaucoma cases [7]) is more challenging due to lack of visible trabecular disorders on a macro scale. In principle OAG could be detected with a finer resolution instrument by directly visualizing, how open are the micrometer-sized trabecular pores. Unfortunately, up to now such a precise instrument was not available and OAG remains largely underdiagnosed in population.

Anticipated micron-resolution instrument would be of high clinical value. With a detailed view of the trabecular meshwork, doctors could determine the source of increased IOP and come up with a treatment particularly suited to this cause. Potential examples include: 1) visible accumulations of fibrillar material blocking the trabecular pores might point towards exfoliation syndrome type of glaucoma with the recommended treatment being outflow-increasing drugs [2], 2) excessive pigment dispersion together with the missing trabecular cells might support a hypothesis of the pigmentary type of glaucoma that is most efficiently treated with prostaglandins medication [8,9], 3) presence of inflammatory debris around trabecular meshwork might suggest an inflammatory glaucoma type that should be treated with inflow (aqueous production) decreasing drugs [2]. A high-resolution instrument would be a promising platform for discovering new in vivo biomarkers associated with each glaucoma type and testing the effectiveness of the new medications. Better understood and treated glaucoma would be of great value for public health as it would improve the outcome for patients and make medicine more targeted: only the patients who face the confirmed risk of developing glaucoma would be required to follow expensive therapies and receive regular medical attention.

Among the instruments that have come the closest to achieving this goal is a gonioscopy device, enhanced via combination with adaptive-optics scanning laser ophthalmoscope (AOSLO) [10]. The images, albeit bringing valuable closer look into the meshwork features, have to be captured with a tilted light beam (about 60º relatively to the optical axis of the eye) to get around the multiply scattering corneosclera overlaying the trabecular meshwork. Such imaging configuration requires physical contact between the optical system and the eye, potentially leading to patient discomfort and making the examination inadvisable for the part of population (children, patients with fragile corneas). Moreover, the instrument lacks the capability of layer-by-layer optical sectioning, potentially creating ambiguity in measurements of the trabecular pores at different depths.

Another class of instruments, based on optical coherence tomography (OCT) [11] demonstrated that looking through an angled view of gonioscopy is not the only possible way to observe the trabecular meshwork. Instead, thanks to the coherent rejection of the multiply scattered photons, OCT can image the meshwork layers directly through the scattering corneosclera [12]. Glaucoma-related studies have so far been focused on exploiting the ability of modern OCT methods, such as spectral-domain OCT (SD-OCT) and swept-source OCT (SS-OCT), to give global views of the entire anterior [13–20] and posterior [21] parts of the eye. In the anterior eye several promising open-angle glaucoma biomarkers were found including the width of trabecular meshwork [22–24] and the volume of Schlemm's canal [25–27].

A different direction was taken by a particular type of OCT called time-domain full-field OCT (TD-FF-OCT) [28,29]. Instead of providing a global view, TD-FF-OCT captures only a small area about 1 mm x 1 mm but with the main advantage of 10× better resolution (about 1 μm) comparing to conventional OCT methods. Given that the pores of ex vivo trabecular meshwork samples are known to also be on the μm scale [30], the device seems particularly suited for exploring this tissue. Unfortunately, in vivo imaging of the meshwork has been so far been set back by the difficulty matching the narrow 1 mm meshwork with 1 mm flat view of TD-FF-OCT (laterally, axially and in depth). More precisely, the existing TD-FF-OCT devices lacked the cross-sectional view that could be used as feedback for locating the region of interest. In principle, TD-FF-OCT could be coupled with SD-OCT to give the additional locating information, however in the known implementation, signal from the SD-OCT was insufficient to recognize any ocular structures except the surface tear film reflection [31]. As such, the device could be used for eye tracking, but not for locating the trabecular meshwork.

In this work we present a novel approach to combine TD-FF-OCT and SD-OCT, which keeps the effective performance of both. The device becomes bimodal, providing two orthogonal views: extended cross-sectional view (for locating the region of interest) via SD-OCT and en face view with μm resolution (for precise diagnostics) via TD-FF-OCT. We explore the potential of the device to facilitate the localization and high-resolution imaging of trabecular meshwork in human eye in vivo. Efforts are taken to make the compact, multi-purpose (cornea, limbus) and low-cost version of the cell-resolution device, advancing its adoption in clinical practice.

## Results

### Instrumentation of orthogonal view OCT

In order to achieve the best performance of both, TD-FF-OCT and SD-OCT were combined in a way that kept them optically independent from each other (sharing the least number of components). This became possible by integrating the two sub-systems through the sample optical arms of their respective interferometers. Thus, the microscope objective and dichroic mirror are the only shared optical parts. Care must be taken to balance the interferometer arms in the presence of the dichroic mirror. More precisely, while the requirements of optical symmetry between the arms are relatively relaxed for spectral-domain methods (optical path length mismatch of a few hundred μm and more is acceptable), the requirements for time-domain methods are strict – sample and reference interferometric arms should be equal in optical path length within the coherence length of the light source (within a few μm). Moreover, time-domain full-field methods, unlike spectral-domain point-scanning methods, are known to be highly sensitive to the wavefront shape symmetry between the optical arms [32,33]. Thus, insertion of a multi-layered dichroic mirror in the sample arm of TD-FF-OCT breaks the optical symmetry through dispersion, optical path length and wavefront mismatches, leading to low contrast and spatial distortion of the interference signal. Typically, one goes around this known problem by inserting the dichroic mirror into the illumination arm [31,34,35], thus keeping the interferometer symmetry intact. However, this severely limits the optical signal and extends the system's footprint substantially. In this work we overcome the problem by inserting two similar dichroic mirrors, one in the sample arm and one in the reference arm, and by performing an additional alignment step (see methods). When aligned, this new configuration produces 4× higher SD-OCT signal comparing to the previous designs, which suffered from the signal loss on both the forward and backward paths going through the 50:50 beam splitter.

Another improvement to SD-OCT image quality comes from the conjugation of the scanning mirror (of SD-OCT) to the microscope objective (of TD-FF-OCT). That conjugation is equivalent to the mirror being placed at the back focal plane of the microscope objective. The system is capable of producing regularly sampled point-by-point illumination beams that are incident perpendicular to the sample surface. Overall, that further increases the signal and improves the quality of SD-OCT images. Moreover, as the SD-OCT is integrated through the dichroic mirror located next to the microscope objective, the 4F conjugation system can be very short (4 × 3 cm = 12 cm), keeping the entire device compact.

Optical and mechanical designs are presented in **Figs. 1a-c**. The whole OCT device is 30 cm × 30 cm × 70 cm (width × longitude × height). These dimensions encompass all the items (TD-FF-OCT, SD-OCT, supporting opto-mechanics, controllers for electronics, joystick stage, etc.) with the only exception being the personal computer located next to the device. Such a small form factor became possible thanks to several solutions. Beyond the aforementioned short conjugation system, the use of a customized commercial compact SD-OCT (OQ LabScope, Lumedica, USA) is the second most important factor in reducing the footprint. Mounting the device on a conventional clinical X-Y-Z axes joystick reduces both the size and cost. Custom fabricated 3D printed and laser-cut parts ensure the tight integration of the components and are keeping the device lightweight.

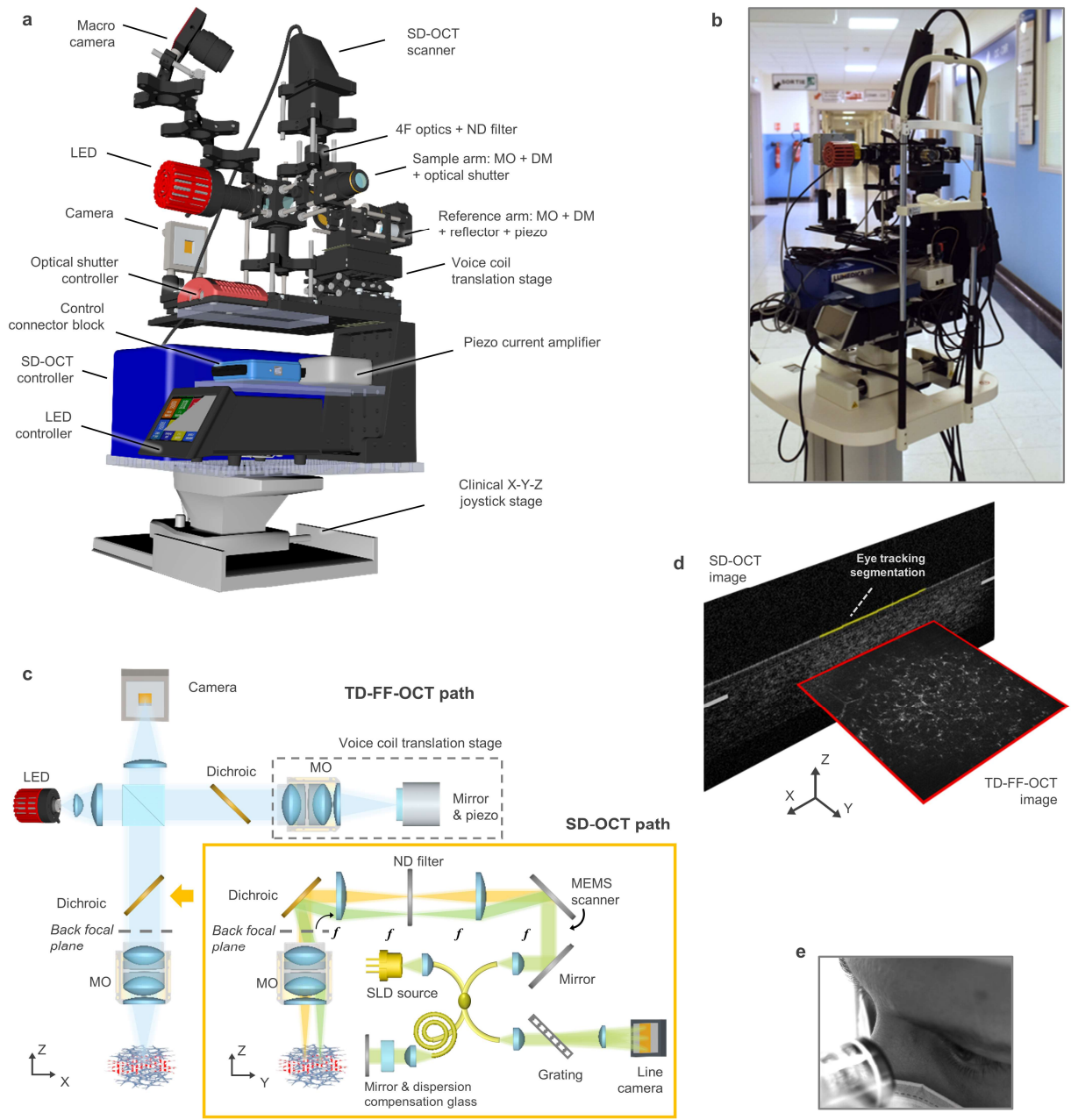

**Fig. 1 | Orthogonal view OCT system. a**, 3D model of the compact prototype. The model encompasses all the main components except the external personal computer. LED - light-emitting diode; SD-OCT - spectral-domain OCT; ND filter - neutral density filter; MO - microscope objective; DM - dichroic mirror. **b**, Photo of the device installed on the clinical table in Quinze-Vingts National Ophthalmology Hospital, Paris. **c**, Optical design. TD-FF-OCT and SD-OCT are integrated through the sample arms of their respective interferometers. MEMS (microelectromechanical system) scanner is conjugated to the back-focal plane of the microscope objective for improving SD-OCT image quality. SLD – superluminescent light diode; Blue lines – 850 nm LED illumination of TD-FF-OCT; Yellow/green lines – 940 nm SLD illumination of SD-OCT at different scanning angles. **d**, SD-OCT (cross-sectional) and TD-FF-OCT (en face) images and their orientations. SD-OCT image is used not only for clinical examination but also for tracking axial eye position (see Methods). **e**, Macro camera view for aligning to the eye.

TD-FF-OCT shows en face 1.2 mm × 1.2 mm (lateral X × lateral Y) views with high 1.7 μm lateral and 7.7 μm axial (in the cornea) resolutions. SD-OCT provides an extended cross-sectional field-of-view (FOV) of 2.3 mm × 3.2 mm (lateral × axial) with resolutions close to that of typical SD-OCT systems: 10 μm lateral and 7 μm axial (in the cornea) (**Fig. 1d**). The two sub-systems are operating in parallel (albeit controlled with separate PCs) and the operator sees the cross-sectional, en face and macro view (**Fig. 1e**) images on the screen. Each sub-system operates at different parts of the optical spectrum: TD-FF-OCT uses 850 nm central wavelength while SD-OCT uses 940 nm. An additional axial eye tracking and correction workflow is implemented to ensure consistent TD-FF-OCT imaging (see Methods).

Each TD-FF-OCT tomographic image is reconstructed from the two phase-shifted camera frames according to the conventional procedure [29,36]. The camera captures the frames rapidly at 550 images/s to suppress eye movements. The tomographic images are displayed live at 10 images/s rate. SD-OCT is sufficiently fast (80,000 A-lines/s) to produce 30 cross-sectional images/s.

**Corneal imaging in healthy subjects and patients**

The study involved two healthy subjects (39 and 41 years old), one elderly volunteer (79 years old) and two clinical patients with ocular pathologies (both 42 years old). The device was practical to be used by the trained orthoptists (see Methods).

In the initial clinical tests, we imaged the central corneal region in healthy volunteers. The combined instrument produced cell-resolution images of similar quality to our former TD-FF-OCT design [31]. We resolved 50 μm superficial epithelial cells, sub-basal nerves and Bowman's layer, stromal keratocyte cells with about 15 μm nuclei as well as endothelial mosaic composed of 20 μm diameter cells (**Figs. 2d-l**). At the same time the quality of the cross-sectional SD-OCT images was much improved. Not only the tear film reflex, but all the corneal layers could be resolved including epithelium, Bowman's layer, stroma and endothelium (**Fig. 2b**).

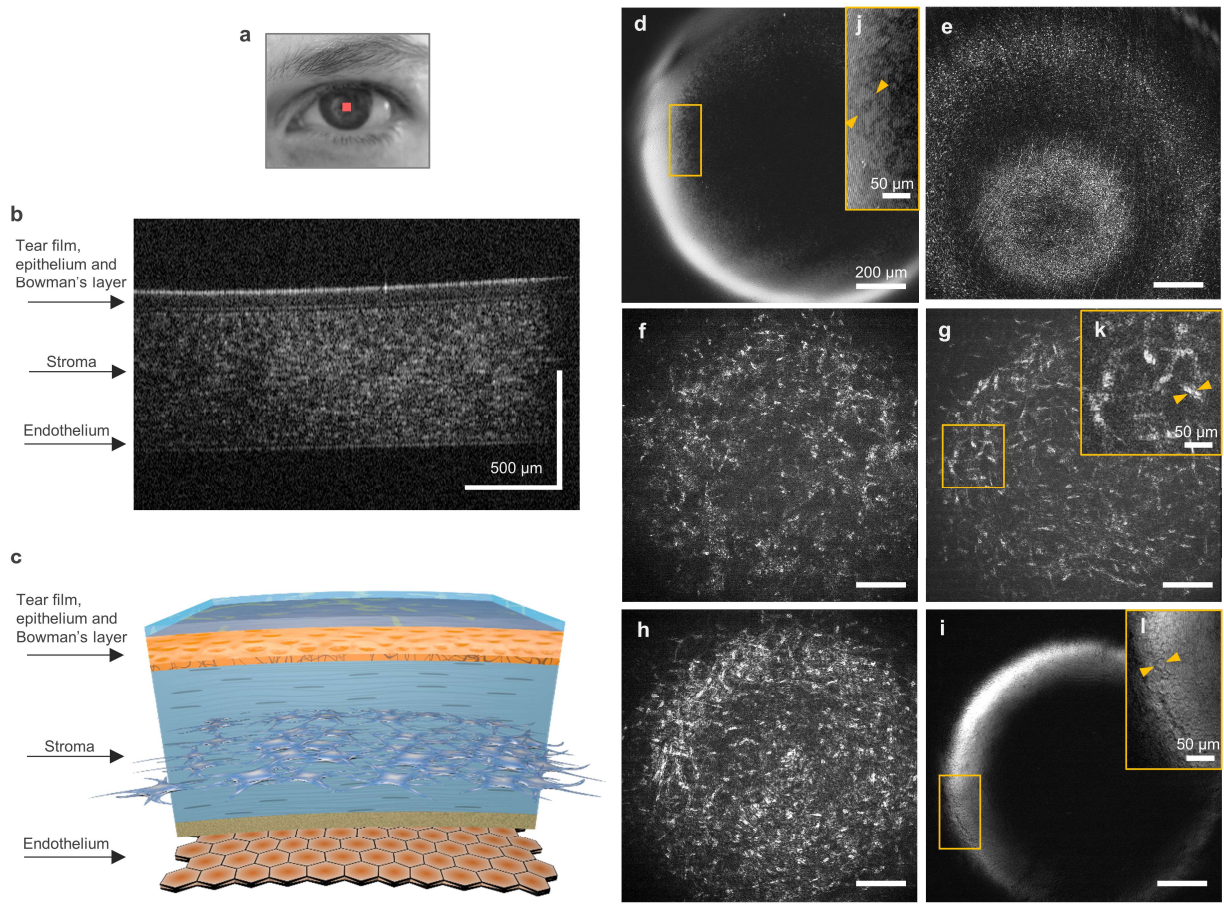

**Fig. 2 | Imaging of healthy human cornea. a**, Images were collected from the central cornea of the heathy subject (red square). **b**, Single (non-averaged) cross-sectional SD-OCT image. The quality was sufficiently high to resolve different corneal layers: tear film, epithelium, Bowman's layer, stroma and endothelium. **c**, 3D representative drawing of the cornea highlighting the layers visible in SD-OCT. **d-l**, TD-FF-OCT en face images of different corneal layers: (**d**) tear film reflex and epithelium with superficial cells seen in inset (**j**), (**e**) sub-basal nerve plexus and Bowman's layer, anterior (**f**), middle (**g**) and posterior (**h**) stroma with visible keratocyte cell nuclei (**k**), (**i**) endothelium formed by hexagonal cell mosaic (**l**). Unlabeled scale bars are 200 µm.

As a next step, we conducted preliminary tests in patients to confirm the feasibility of clinical use and to determine the most promising directions for the future clinical research. The cell-resolution TD-FF-OCT images were of central interest. The explored ocular conditions included age-related corneal changes, keratoconus and Fuchs' endothelial dystrophy.

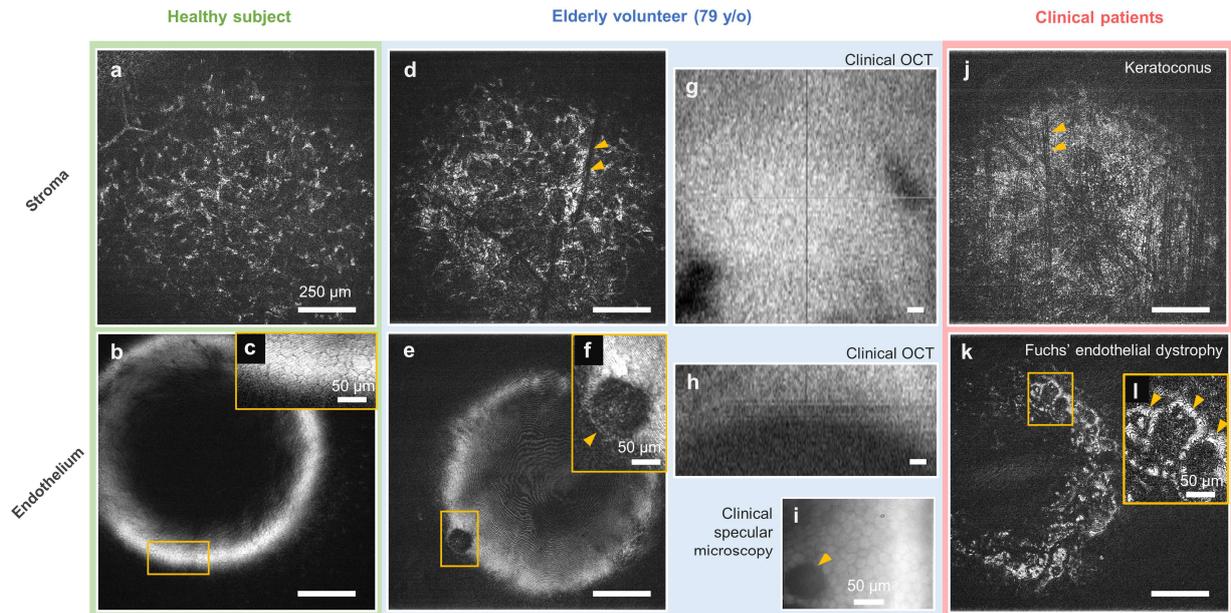

**Fig. 3 | High-resolution corneal imaging in clinical patients. a-c**, Reference images of healthy human corneal layers as seen by TD-FF-OCT. (**a**) stroma; (**b**) endothelium with visible endothelial cell mosaic (**c**). **d-f**, TD-FF-OCT images from an elderly volunteer. Basic clinical (slit-lamp) exam preceding the high-resolution imaging has found no signs of ocular pathologies. On the contrary, TD-FF-OCT has easily identified abnormalities in the stroma (**d**) and endothelium (**e**). Yellow arrows in the stroma point at dark bands – striae, indicative of corneal mechanical stress. In the endothelium the arrow points at dark spot – guttata (**f**), the outgrowth of Descemet's membrane. **g-i**, Comparison en face images from the same elderly subject reconstructed by the available clinical modalities: stroma (**g**) and endothelium (**h**) views with anterior eye OCT (RTVue XR 100, Optovue, USA); endothelium view (**i**) with specular microscopy (SP-3000P, Topcon, Japan). **j-l**, Preliminary TD-FF-OCT images from the clinical patients. Cornea of keratoconus patient (**j**) had bright scattering background and was fenestrated with the dark striae highlighted by the yellow arrows. Fuchs' endothelial dystrophy patient had severely altered endothelium (**k**) covered with guttata spots (**l**). Unlabeled scale bars are 250 µm.

The elderly subject showed dark bands 5–30 µm wide passing through the corneal stroma (**Fig. 3d**). We hypothesize those to be stromal striae - tissue folds indicative of corneal mechanical stress [37]. Moreover, the stromal background was significantly more reflective in this subject comparing to the healthy volunteer. This might be due to the disorganization of stromal fibers that causes increased backscattering seen as bright haze. At the endothelial corneal region, infrequent 50–100 µm wide dark guttata spots (the outgrowths of Descemet's membrane (**Fig. 3e,f**)) could be seen. Presence of guttata was confirmed with the clinical high-resolution specular microscope (SP-3000P, Topcon, Japan) (**Fig. 3i**). Comparison between the TD-FF-OCT images and the reconstructed en face images from a clinical SD-OCT (RTVue XR 100, Optovue, USA) (**Fig. 3g,h**) highlights the difference in benefits associated with using each device. While the large viewing area of clinical SD-OCT is best suited for global exploration of tissue for pathologies, cell-resolution of TD-FF-OCT is particularly adapted for evaluating the local cellular organization. Importantly, cell organization is indicative of the general tissue properties (such as optical transparency, mechanical stress) that have an effect over the entire cornea.

Next, we imaged a patient with keratoconus, an ocular condition that alters the corneal shape (**Fig. 3j**). Deformation of the cornea led to the appearance of stromal striae that were more numerous compared to the elderly subject and were predominantly vertically oriented (Vogt's striae) rather than criss-crossed due to the deformation of the cornea and the straining toward the conical apex. The striae were visible as 5 μm black stripes organized into 10-100 μm thick bundles. Furthermore, disorganization of the corneal structure happened not only at micron scale. Bright stromal background (stromal haze) indicated increased back-scattering from the misaligned nanometric stromal fibrils. Compared to the elderly subject this back-scattering was more significant: no bright keratocyte cells could be identified against the bright background.

A patient with Fuchs' endothelial dystrophy showed a severely affected endothelium (**Fig. 3k**): the layer was densely covered with guttata (**Fig. 3l**), while the remaining parts of the cell mosaic were rarely visible.

**Micrometer-level structure of ex vivo trabecular meshwork**

Before imaging trabecular meshwork in humans (in vivo), we were interested to learn how these structures look (ex vivo) under the high-resolution OCT views. We used a commercial ex vivo immersion-oil TD-FF-OCT device (**Fig. 4a,b**) with isotropic 1 μm resolution. The human anterior eye from the tissue bank (see Methods) was dissected from the crystalline lens and flipped so that the corneal endothelium was facing the microscope objective (**Fig. 4c**). In the limbal region the trabecular meshwork fibers could be immediately seen (**Fig. 4d**). The cross-sectional views (**Fig. 4e**) showed characteristic lamellae sheets separated by the intertrabecular openings and interlamellar spaces well known from the literature [38].

Anterior trabecular meshwork interfacing the corneal endothelium was of primary interest. On the surface the fibers measured 2-10 μm thick and were oriented predominantly parallel to the corneal-limbal boundary (**Fig. 4f**). The fibers formed pores about 15 to 70 μm in diameter, as measured in the single meshwork layer (**Fig. 4g**). This range might represent the lower bound of the possible diameters because the anterior meshwork was compressed and tilted (as can be seen in **Fig. 4e**). The location and sizes of the above structures resembled trabecular meshwork of uveal type, known from the ex vivo literature [30,39,40]. We could also notice the bright oval-shaped spots about 10 µm matching in appearance to the nuclei of the trabecular meshwork cells [39].

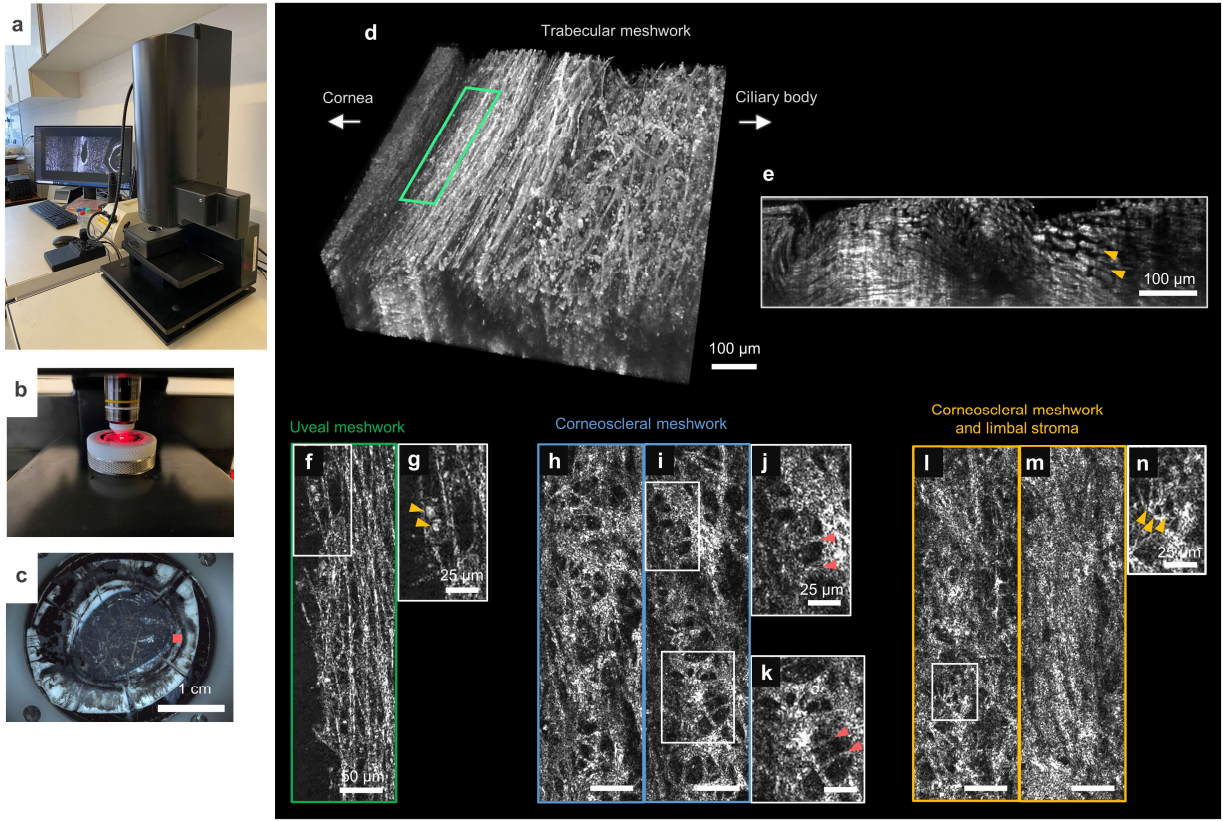

**Fig. 4│ Ex vivo trabecular meshwork in tomographic views. a-c**, Ex vivo TD-FF-OCT device (LightCT scanner, Aquyre Biosciences, USA (former LLTech, France)) (**a**) with a close-up view on the sample holder (**b**) and anterior eye sample (flipped with corneal endothelium facing the microscope objective, crystalline lens being removed) (**c**). Red rectangle shows approximate imaging location in the posterior limbus. **d**, 3D-view of the trabecular meshwork. Green rectangle corresponds to the anterior meshwork region presented in (**f-n**). **e**, Cross-sectional slice of the volume (**d**). Trabecular lamellae sheets are clearly separated (yellow arrows). **f,g**, Uveal meshwork with pores elongated along the corneal-limbal boundary and trabecular cell nuclei (yellow arrows). **h-k**, Corneoscleral meshwork (20-30 µm deep axially) with finer pores and fibers. Fibers are mainly protruding in two orthogonal directions (along the corneal-limbal boundary and perpendicular to it (red arrows)). **l-n**, Corneoscleral meshwork (40 µm deep axially) and limbal stroma (50 µm deep axially). With increasing depth, the pores and fibers (yellow arrows) are shrinking in size down to the resolving limit of the optical system (1 µm). Unlabeled scale bars are 50 µm.

At the deeper layers (axially 20-30 µm from the endothelium towards the ocular surface), the meshwork became denser appearing as perforated sheets (**Fig. 4h,i**). Comparing to the uveal meshwork the pores were smaller about 5 to 30 µm in diameter. The pores were formed by the thick 10-20 µm trabecular beams. Internally the beams had finer fibers measuring 2-6 µm in diameter (**Fig. 4j,k**). The fibers were predominantly parallel or orthogonal to each other with preferred orientations being towards the cornea or along the corneal-limbal boundary. The mentioned meshwork structures share many similarities with corneoscleral meshwork, studied in ex vivo literature [30,39].

Finally, with further increasing depth (axially 30–50 μm) the pores and fibers were decreasing in size down to the resolution of the optical system (**Fig. 4l-n**).

**Imaging human trabecular meshwork in vivo**

In order to access in vivo trabecular meshwork at the base of the limbus, one needs to pass through the 1 mm-thick corneoscleral tissue. Significant multiple scattering of this tissue prohibits meshwork imaging with conventional microscopy, unless using an angled gonioscopic view through the transparent cornea. On the other hand, OCT techniques are known for their ability to select ballistic single-backscattering photons in presence of multiple-scattering background. This motivated us to attempt direct imaging through the corneosclera (**Fig. 5a,b**).

The cross-sectional view of SD-OCT was crucial for the fine alignment to the trabecular meshwork. First, the anterior eye angle was located in the SD-OCT image. Using the image as feedback, the subject was asked to rotate the eye until the limbal cornea was oriented perpendicular to the incident light (**Fig. 5d**). When aligned, the bright Schwalbe's line corresponding to the termination of endothelium and the beginning of trabecular meshwork became clearly visible. The entire device was moved to align the Schwalbe's line to the center of the SD-OCT imaging field, which was matched with the TD-FF-OCT field by optical design.

Trabecular meshwork structures seen with TD-FF-OCT were varying with depth. At the deep interface with aqueous humour we saw the bright endothelium and meshwork-like structures (**Fig. 5o-r**). The endothelial layer was covered with interference fringe artifacts, which are commonly seen in smooth layers by TD-FF-OCT [41]. The endothelium terminated at irregular lateral locations within the 400 μm transition zone, previously reported in electron microscopy studies [38,42].

The meshwork was visible as loose connective tissue beams that formed predominantly oval openings. The oval openings were extended in the direction along the Schwalbe's line and the beams were lying in a single en face plane. The beams were about 10-30 μm thick while the openings showed great variation in sizes - 70-150 μm along the long and 30-60 μm along the short oval axes. These values resemble those of ex vivo trabecular meshwork of uveal type seen in the literature [30,39,40].

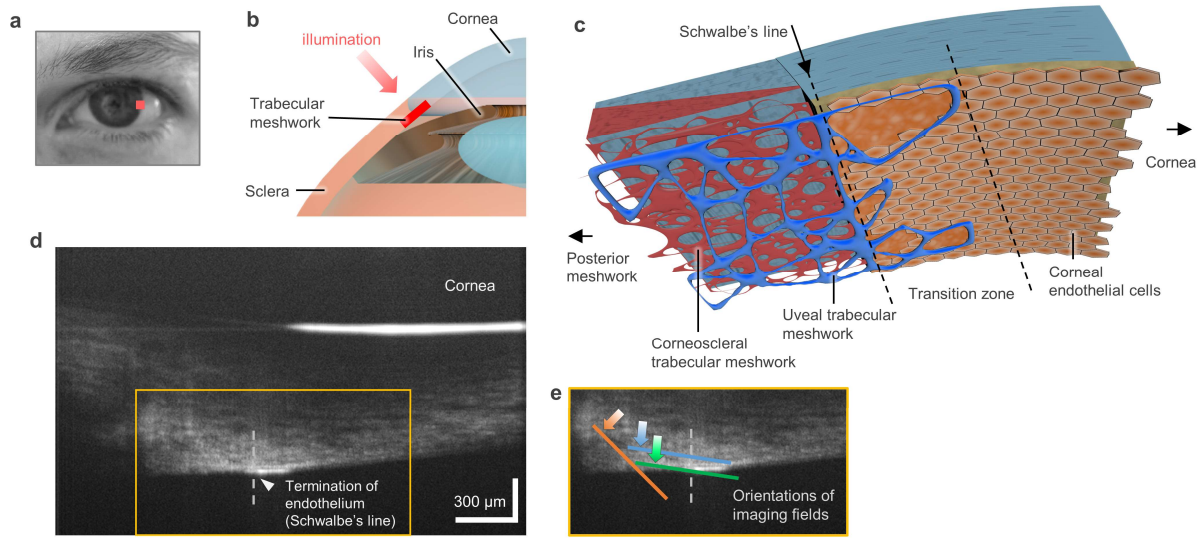
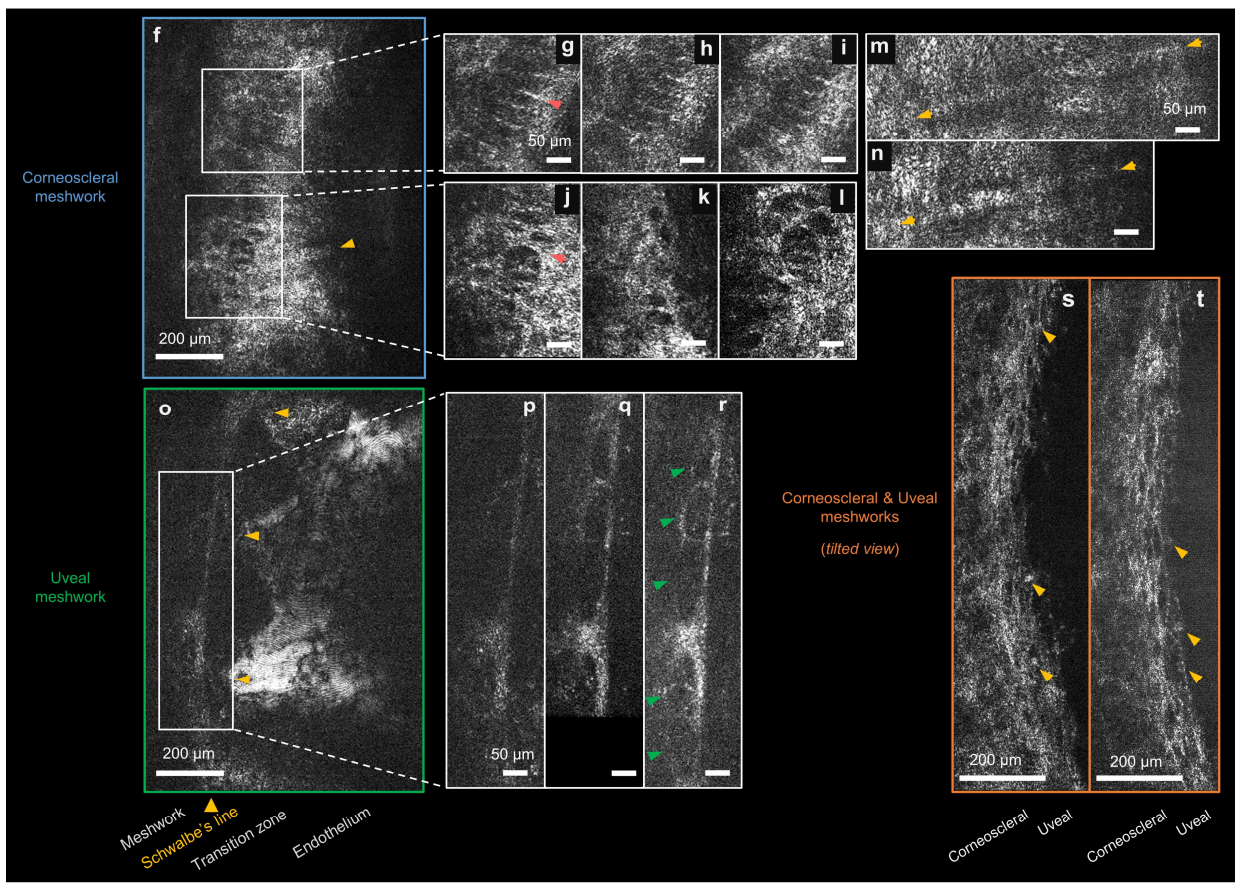

**Fig. 5 | In vivo human trabecular meshwork imaging. a**, Macro eye view showing in red the approximate imaging location (temporal limbus of the left eye). **b**, Orientation of illumination relatively to the trabecular meshwork in the anterior eye. Thanks to the ability to select the ballistic scattering photons, TD-FF-OCT can partially image through the scattering corneosclera. **c**, 3D representative image of the cornea-trabecular meshwork interface. **d**, SD-OCT image highlighting the bright termination of endothelium and the beginning of trabecular meshwork. This image was crucial for aligning to the meshwork. **e**, Zoom of image (**d**) showing the orientation of en face (blue and green) and tilted (orange) imaging

configurations. **f**, Corneoscleral trabecular meshwork formed by the porous sheets (**j-l**) and fibers at the core of the beams (red arrows) (**g-i**). The same area also hosts the thick and long ciliary body tendons (yellow arrows) (**f, m, n**). **o-r**, Uveal trabecular meshwork composed of loose fiber network and wide pores (green arrows). Endothelium is attached to the trabecular meshwork at irregular locations (yellow arrows). **s,t**, Trabecular meshwork layers seen in tilted views. Transition from the wide (uveal) to small (corneoscleral) pores is seen from the right to the left of the image. Yellow arrows point at oval particles resembling the sizes expected from the trabecular cell nuclei. Unlabeled scale bars are 50 µm.

Above the uveal layer we saw the meshwork of another type visible as perforated sheets (**Fig. 5f-l**). The holes were oval, elongated in the direction perpendicular to the Schwalbe's line, and smaller about 40 µm along the long and 15 µm along the short oval axes. The beams were about 15 µm thick, with each beam being formed as a sheath around a single fiber about 7 µm thick. This internal structure was occasionally revealed, when the optical coherence plane happened to be sectioning through the beam. The fibers of each beam were parallel protruding in two preferential orientations - towards or perpendicular to the corneal direction. The spacing between the fibers was regular about 20 µm. These meshwork structures appear similar to the corneoscleral meshwork, studied above and in ex vivo literature [30,39]. Within the same meshwork we also noticed thick (30 µm) and long (length above 700 µm) straight bands (**Fig. 5f,m,n**). We hypothesize those to be the ciliary body tendons that are known to traverse the trabecular meshwork and insert into the posterior corneal stroma [43–45]. Tendons were oriented orthogonally to the Schwalbe's line.

The images presented above were collected from the region of trabecular meshwork adjacent to the cornea. Using the same optical configuration, we attempted to image the entire meshwork. However, the influence of multiple scattering from the sclera was stronger in farther meshwork regions, prohibiting imaging. In order to access the larger meshwork portion, we tilted the optical axis so that both the illuminated and detected light beams would pass through the transparent cornea. As a result of the tilt each tomographic image sectioned through several meshwork layers (**Fig. 5s,t**). The visible sizes of the pores were varying from 80 µm on the right side of the tissue (uveal) through 30 µm in the center (corneoscleral) to 8 µm on the left. Although, we have noticed even smaller 2 µm pores, we could not confidently resolve those due to speckle-like appearance of the tissue. Finally, in the tilted views we saw the bright oval spots about 10-15 µm as expected for the nuclei of ex vivo trabecular meshwork cells [39].

# Discussion

Open-angle glaucoma presents a complicated problem for the current medical system because the disease evolves slowly without noticeable symptoms. As a result, patients present themselves to the ophthalmic service at late disease stages, when a significant part of the visual field has already been irreversibly lost. In these circumstances the major responsibility to constrain the disease lies on the preemptive general population screening programs. However, at present the decision to establish these programs has been widely turned down due limited accuracy of the early-glaucoma screening tests [46,47]. Inaccurate overdiagnosis and overtreatment can lead to harmful side effects in mass population (eye irritation, surgical complications, cataract, etc.) outweighing the potential benefits of screening. One can potentially improve screening accuracy by adopting the advanced retinal and optic nerve imagers into the future clinical practice. Indeed, modern adaptive optics systems can visualize retina with much higher detail on a level of single nerve fibers and ganglion cells [48–50], polarization-sensitive OCT systems are sensitive to small changes in density of ganglion cell axons or their microtubules [51], OCT angiography [52] can detect micrometer alterations in vessel structures while Doppler holography [53,54] measures millisecond changes in blood flow, full-field swept-source OCT shows nanometric changes of ganglion cells thickness following the functional light stimulus [55], TD-FF-OCT can view micrometer-size nerve fibers using the simplified optical design thanks to relative insensitivity to optical aberrations [35,56], etc. A fundamental challenge for retinal imaging devices is that they have to operate within a short timeframe of early glaucomatous damage. More precisely, these devices are not able to detect the pressure build-up in the eye over time and need to wait until the smallest signs of retinal damage appear. Naturally, these smallest signs are difficult to detect with high accuracy.

In an alternative approach one could try to detect not the consequence but the typical source cause of glaucoma – blockage of trabecular meshwork pores. The latter precedes the ocular pressure elevation and any posterior eye damage; therefore, structure of trabecular meshwork may show significant deterioration even in early glaucoma. The desired screening instrument should provide 1) sufficiently high resolution to resolve the openness of trabecular meshwork pores, 2) en face view that matches in orientation with trabecular layers and optical sectioning to clearly show the pores, 3) rejection of multiple scattering photons from the corneosclera. In this work we have created such a device using a combination of TD-FF-OCT and SD-OCT optical techniques.

We foresee a high clinical value for cell-resolution images of trabecular meshwork. In particular, openness and shape of pores is one biomarker that would be promising to investigate in the first place. For example, shrinkage of pores in direction parallel to Schwalbe's line could suggest a direct way of increasing outflow via ciliary muscle contracting drugs that spread the pore spaces open [57]. Visualization of ciliary muscle tendons and measurement of their thickness could tell how efficiently the contraction forces spread across the meshwork. Reduction of pore sizes in all directions could suggest the increase in sheath thickness surrounding the elastic fibers. From the ex vivo studies the sheath increase is known to be associated with aging and open-angle glaucoma [58]. A potentially promising possibility of exploring the internal composition of the trabecular beam (sheath/elastic fiber) is available thanks to optical sectioning. Reduction in visible trabecular cell nuclei could be an indicator to initiate therapy. More precisely, trabecular cells are responsible for eliminating the constantly accumulating debris in the meshwork – the critical role that restricts the uncontrollable growth of IOP [59]. The new instrument could potentially determine the type of particulate matter blocking the trabecular pores: inflammatory debris, accumulations of white blood cells, extracellular deposits, lysed red blood cells, or other solid components. The detected material would give evidence about the open-glaucoma class - primary/secondary and type – steroid-induced, pigmentary, pseudoexfolliative, uveitic, etc. The precise diagnosis could improve the outcome for the patients as doctors would use the treatment adapted to the particular glaucoma type.

Orthogonal view OCT would also be an invaluable platform for understanding the cellular-scale biological processes that regulate the aqueous flow in vivo. Another interest is classifying the efficiency of existing and emerging therapies as well as the efficiency of classical and minimally invasive surgical interventions targeting the trabecular meshwork.

Higher diagnostic accuracy could open a path to the future adoption of preemptive general population screening programs. Successful programs could have a positive effect both on society through improving the ocular health in the population and on economy through reducing the number of costly surgical interventions and through keeping the population economically active.

In order to become a practical reality the device must not only detect an important clinical problem, but must be compact enough to fit in a scarce clinical spaces, be inexpensive, allow rapid testing, and ideally be operatable by the mid-level or orthoptist-level users [2]. The design of orthogonal view OCT was an attempt to comply with these requirements. The effective cost was reduced by implementing several imaging capabilities in one instrument enabling observation of: trabecular meshwork, cornea with cell-detail (partly equivalent to confocal microscopy), corneal endothelium with cell-detail (partly equivalent to specular microscopy) and cornea/limbus in cross-section (partly equivalent to anterior eye OCT). In the cornea, the device produces images comparable to the other ex vivo [60,61] and in vivo [62–66] high-resolution corneal imaging research methods, while providing benefits of contactless usage and 10× larger FOV comparing to the clinical state-of-the-art [67]. The first clinical results highlight the key benefit of high-resolution imaging – access to the local tissue properties such as optical transparency and mechanical stress that have an effect over the entire cornea. Moreover, being non-contact, the device can directly correlate the striae bands with the internal mechanical stress in the cornea, which is challenging with the existing contact devices that may themselves induce the mechanical stress. Future large scale clinical trials would be important for finding novel biomarkers that could explain and predict the progression of the diseases on the cellular level.

Development of next generations of OCT devices capable of more efficient rejection of multiply scattered photons would be promising for accessing the entire aqueous flow pathway (trabecular meshwork, Schlemm's canal, and aqueous veins). Another direction of great interest would be to supplement the scattering contract of OCT with the functional information on aqueous flow or mechanical stiffness measurement in the meshwork [68–70], the latter being among the key glaucoma biomarkers.

**Methods**

**Orthogonal view OCT device (SD-OCT sub-device)**

We used a customized version of the commercial SD-OCT device (OQ Labscope, Lumedica, USA). The original device relies on a classical optical SD-OCT principle, however implements several unique solutions to keep the cost and size low [71]. Customization included the new superluminescent light diode (SLD) that had a 940 nm central wavelength, avoiding the overlap with 850 nm light of TD-FF-OCT. The axial resolution of 7 μm in the cornea was estimated from the spectral bandwidth of 40 nm, which was experimentally confirmed with the spectrometer (CCS175/M, Thorlabs, USA). In order to integrate the device into the TD-FF-OCT we incorporated the dichroic mirror (DMSP900, Thorlabs, USA) into the sample arm of SD-OCT interferometer. We have also added a lens (LA1805-B, Thorlabs, USA) that in combination with the original lens of SD-OCT (LA1289-B, Thorlabs, USA) creates the 4F optical system. The 4F system conjugates the scanning mirror of SD-OCT to the back-focal plane of TD-FF-OCT microscope objective. The SD-OCT beam fills only a small portion of objective's numerical aperture (NA) leading to reduced lateral resolution of 10 μm (verified with the resolution target). At the same time, the lower NA plays a positive role by increasing the depth-of-focus so that the entire thickness of the cornea can be seen in the image. The above extension of the sample interferometer arm needs to be compensated in the reference arm. The reference optical fiber was extracted from the system and set on a rail enabling the precise control of the reference arm length. Blocks of glass (LSM03DC, Thorlabs, USA) are inserted in the arm to compensate for the glass of microscope objective and lenses in the sample arm. In order to stay within the maximum permissible exposure of ISO 15004-2:2007 light safety standard, we reduced the power coming out of SD-OCT to 0.4 mW using the neutral density filter (NENIR02A, Thorlabs, USA). The SD-OCT scanner head is mounted on a custom-made 3D printed holder. SD-OCT acquires images at a rate of 80,000 A-lines/s. Display rate is 30 B-scans/s. Original C# code was customized to display the surface segmentation in real-time (see below).

**Orthogonal view OCT device (TD-FF-OCT sub-device)**

TD-FF-OCT is a Linnik interferometer, equipped with the two identical microscope objectives, a 2D camera and a light-emitting diode (LED) source of low spatial and temporal coherence. Microscope objectives (LMPLN10XIR, Olympus, Japan) and a 250 mm tube lens (AC254-250-B, Thorlabs, USA) together produce 13× magnified image on an about centimeter-size sensor resulting in 1.2 mm field of view. Numerical aperture of 0.3 results in 1.7 μm lateral resolution that was theoretically estimated according to the Rayleigh criterion and experimentally confirmed with a resolution test target as well as by measuring the diameters (FWHM) of 80 nm gold nanoparticles imaged on a glass plate. The working distance of the objectives is 18 mm, sufficient to avoid the risk of accidental physical contact with the eye. Illumination is provided by a near-infrared (NIR) 850 nm light-emitting diode (LED) source (M850LP1, Thorlabs, USA). The LED is controlled by the high-power driver (DC2200, Thorlabs, USA). The axial resolution of 7.7 μm in the cornea was estimated from the experimentally measured spectral bandwidth of the LED (30 nm) with the spectrometer (CCS175/M, Thorlabs, USA) and by using the known average corneal refractive index of 1.376 [72]. Light from the LED is collimated by an aspheric condenser lens (ACL12708U-B, Thorlabs, USA). Before entering the objective, light from the source is equally separated by the 50:50 beam splitter cube (BS) (BS014, Thorlabs, USA) into the sample and reference arms of the interferometer. The objective in the reference arm focuses light onto an absorptive neutral density (ND) glass filter (NENIR550B, Thorlabs, USA), which plays a role of a single mirrored surface with 4% reflectivity. Use of an ND filter instead of a glass plate enables elimination of ghost reflections from the back surface of the filter. A low reflectivity value is chosen to achieve high detection sensitivity, which is maximized when the total reflectivity (from all the layers participating or not to the TD-FF-OCT signal) of the sample and reflectivity of the reference mirror match, as indicated by the known TD-FF-OCT signal-to-noise calculations [36]. Total reflectivity from all ocular surface layers, estimated from the Fresnel relations, is around 2%. By using a reference mirror with a reflectivity of 4%, we can expect sensitivity close to the ideal condition. Light in the sample arm, backscattered from the different planes in the eye, and light in the reference arm, backscattered from a single mirror plane, are collected by the microscope objectives and get recombined on the BS. This results in interference, but only for the light coming from a single sample plane and light coming from the reference mirror plane, which match in terms of optical path length. The temporal coherence length of the light source (associated with spectral bandwidth) determines the thickness of the interference fringe axial extension and therefore the optical sectioning precision, which is 7.7 μm. Interfering light and non-interfering light, arising from the other planes of the eye, are focused onto a sensor by a magnifying tube lens (AC254-250-B, Thorlabs, USA). The sensor is a high-full well capacity (2Me-) 1440×1440 pixel CMOS camera (Q-2A750-CXP, Adimec, Netherlands), which captures each 2D image in 1.75 ms. In order to perform tomography, the light that has interfered must be extracted from the background. The useful signal is retrieved by a two-phase shifting scheme [36], where the position of the reference mirror is rapidly modulated using the piezo mirror-shifter (STr-25/150/6, Piezomechanik GmbH, Germany) synchronized with the camera. In order to increase the piezo motor thrust and acceleration, it is supplied with an increased current from a custom-made current amplifier. The camera captures two consecutive images that are equivalent with the only exception being the phase of the light coming from the shifted reference mirror. Between the two frames the reference mirror is shifted by about 200 nm, thus providing the π phase shift (accounting for the double pass). By subtracting the two images and taking an absolute value we remove the non-interfering signal (that is not affected by the phase shift) and double the interfering signal, thus producing a tomographic image. The two camera frames are captured rapidly at 550 frames-per-second (fps) to suppress the eye movements. It takes 3.5 ms to capture two camera frames (one tomographic image). Image acquisition and real-time display code is written in Labview 2014 (National instruments, USA). Labview orchestrates the images acquisition via the input/output device (PCIe-6353, National Instruments, USA) and control connector block (SCB-68A, National Instruments, USA).

To ensure ocular safety according to the European ISO 15004-2:2007 standard, we used pulsed light. One light pulse had duration of 3.5 ms, sufficiently long to rapidly acquire a tomographic image. The break between the light pulses was 100 ms, resulting in 10 fps display and acquisition rates. It should be noted that the most recent US ocular safety standard ANSI Z80.36-2016 imposes a much less strict limitation for exposure of the anterior eye, which in the future potentially enables use of our device without LED pulsation in the US.

A complete evaluation of corneal and retinal safety was performed similarly as detailed in our previous work [31]. Given that the light is sharply focused by the microscope objective on the cornea, light becomes widely spread on the retina. Thus, corneal safety determines the light exposure limit. The results are summarized below. We took into consideration the durations of pulsed light exposures (3.5 ms for LED and 2.78 ms for SLD (beam within 1 mm averaging aperture during scanning)), durations of breaks without the light (100 ms for LED and 3.6 ms for SLD) and irradiances per pulse (2 W/cm$^2$ for LED and 40 mW/cm$^2$ for SLD). These values were used to calculate the corneal exposure. The calculated time-dependent graphs show that during real-time acquisition we reach only 36% of maximal permissible exposure (according to ISO 15004-2:2007).

In order to integrate SD-OCT into the TD-FF-OCT we inserted a dichroic mirror (DMSP900, Thorlabs, USA) in the sample arm of TD-FF-OCT. Unfortunately, the insertion of the dichroic mirror breaks the optical symmetry leading to low contrast and spatial distortion of the interference signal. As a solution to this problem, we inserted two similar dichroic mirrors, one in the sample arm and one in the reference arm. Then we performed an additional alignment step. At first, both dichroic mirrors were oriented perpendicular to the incident light. The TD-FF-OCT was aligned in the conventional way and showed characteristic interference patterns. As soon as we started to tilt one dichroic mirror, the interference signal became distorted and decreased in contrast. By gradually tilting in the same direction the second mirror we could see the recovery of the signal. Step-by-step, we could tilt both mirrors until they became oriented at 45 degrees to the incoming light. The final measured interference signal was as high as in conventional TD-FF-OCT without dichroic mirrors.

**Eye tracking and compensation of optical mismatch**

As was mentioned above, TD-FF-OCT is highly sensitive to the optical symmetry of the interferometer. In fact even introduction of the sample into the interferometer shifts the focus of the light beam (due to Snell–Descartes law) and breaks this symmetry [73]. More precisely, when the light is focused on the surface of the sample, both arms of TD-FF-OCT interferometer are matched. However, as the sample shifts axially, the interferometer arms become mismatched due two factors: 1) the spread of the focal point into the sample due to Snell's law, originating from the large illumination angle and difference between the refractive indexes of ocular surface (1.376) and air (1.0), 2) the shift of the coherence plane in the opposite direction towards the objectives, also caused by the refractive index difference. This known problem is referred to as defocus in TD-FF-OCT [73]. If the axial position of the sample is known, one can compute the focus position and determine the extension of the reference arm required to bring interferometer back to the optical balance. The difficulty is that the live human eye is constantly involuntarily moving [41], thus axial tracking and correction should be performed in real-time. In our former work, we have shown that the tear film reflex detected by the SD-OCT can be used for eye tracking [31]. The difference with the present device is that the quality of the SD-OCT image has been much improved – not only the tear film but all the corneal layers became visible. Segmenting corneal surface in the presence of many reflective layers is more sophisticated than detecting a single axial peak as was done previously. Therefore, we wrote a new custom algorithm for fast (6 ms) curved surface segmentation (see Supplementary code).

First, the algorithm detects the peak signal within each A-line. Then, it checks for existence of other peak signals close to the main peak in intensity, but located closer to the top of A-line (closer to the surface). The peak curve across the A-lines is smoothed to remove outliers. Finally, the algorithm looks at the local smoothness of the final curve (from one A-line to the next), removes the segments that had larger curvature than expected in the ocular surface and keeps the largest clean segment. Segmentation took 6 ms on average as measured by the performance estimator of Matlab R2020a (Mathworks, USA). The Matlab code was converted to .NET assembly and implemented into the C# code of SD-OCT. The segmentation was displayed on top of the SD-OCT B-scan image in real-time.

The algorithm detects the location of the cornea to within the precision of the axial resolution (7 µm). SD-OCT is sufficiently fast (80,000 A-lines/s) to produce 30 cross-sectional images/s with each image being used to detect a new axial eye position. The detected surface location was fed into the voice-coil translation stage (X-DMQ12P-DE52, Zaber, Canada) to keep the TD-FF-OCT interferometric arms optically matched in real-time. The motor was pre-set for 2 µm accuracy, 1 mm/s velocity and the smallest 1 ms latency on the communication port. For faster communication SD-OCT and translation stage of TD-FF-OCT were driven by the same C# software on the same personal computer. Aside from the tracking and correction, independent personal computers were used for image acquisition in SD-OCT and TD-FF-OCT.

**Clinical device design**

For additional safety we implemented an optical shutter with controller (SHB05, Thorlabs, USA) that provides means for blocking the light independently from the software. We also added a macro camera (CS165MU/M, Thorlabs, USA) for easy alignment to the eye. We manufactured numerous custom parts to integrate the entire instrument (except the personal computer) into the compact 30 cm x 30 cm x 70 cm volume. Small parts were designed with Blender open-source software and were 3D printed on Prusa MK3S+ (Prague, Czech Republic), while the large 2D parts were designed in open-source Inkscape and cut with Trotec Engraver Speedy 100 (Trotec, Germany) laser cutter. The entire device was mounted on a clinical table with a standard manual X-Y-Z joystick (Imagine Eyes, Paris, France).

**Ex vivo imaging**

Human donor cornea (with limbus) was obtained from the tissue bank of the Etablissement Français du Sang, Ile-de-France (Paris, France) after it had been rejected for transplantation due to low endothelial cell count. The study was carried out according to the tenets of the Declaration of Helsinki and followed international ethical requirements for human tissues. The sample was preserved in CorneaMax (EuroBio, France) medium for a maximum of 35 days at 31 °C, in accordance with European Eye Banking regulations. It was then placed in CorneaJet (EuroBio, France) medium containing Dextran for deturgescence 48 hours prior to imaging, and transferred to the Quinze-Vingts National Ophthalmology Hospital (Paris, France).

For imaging we used a commercial ex vivo TD-FF-OCT device (LightCT scanner, Aquyre Biosciences, USA (former LLTech, France)). The device utilized a broadband white-light halogen source (filtered for red and near-infrared) and 0.3 NA oil-immersion microscope objective to provide high isotropic $1 \times 1 \times 1$ µm³ resolution. The field-of-view was 800 µm. The sample was placed in the holder and pressed gently against the glass coverslip.

**In vivo imaging**

Prior to the study, all healthy subjects and patients underwent a routine clinical exam that confirmed their ocular condition. Informed consent was obtained from all volunteers and the experimental procedures adhered to the tenets of the Declaration of Helsinki. Approval of the study was obtained from the CPP (Comité de Protection de Personnes) Sud-Est III de Bron and ANSM (Agence Nationale de Sécurité du Médicament et des Produits de Santé), with corresponding study numbers 2019-021B and 2019-A00942-55. The study was carried out at the investigation center of the Quinze-Vingts National Eye Hospital. OCT was mounted on a standard clinical joystick allowing alignment to the subject's eye in 3 dimensions. The joystick was placed on a standard clinical table capable of motorized vertical alignment for the patient's comfort.

The subjects were asked to rest their chin and forehead on a standard headrest while fixating on a target. Examination was non-contact and no medication was introduced into the eye. Light illumination, visible as a dim red circular background, was comfortable for viewing, due to the low sensitivity of the retina to NIR light. Exposures were within the limits specified by the up-to-date ISO 15004-2:2007 standard used in France. While one eye was imaged, the fellow eye was fixating on a target. When imaging non-central parts of the eye, the subject's head was tilted by the examiner to position the plane of ocular surface perpendicular to the direction of the incoming light beam. Total corneal examination was fast (a couple of minutes to align and acquire a stack of images), however trabecular meshwork imaging required substantial time spent on aligning the subject's limbus perpendicular to the beam in the absence of the non-central focusing target.

The instrument was used by an orthoptist, who underwent a 1-week device-usage training. The examiner could precisely line up the device to the eye using the macro camera view (**Fig. 1e**). When aligned, the display showed the en face TD-FF-OCT and cross-sectional SD-OCT images. For easy navigation in the eye, the SD-OCT image was overlaid with the corneal surface segmentation as well as with the line indicating the approximate depth, at which the TD-FF-OCT images were being captured. TD-FF-OCT images had the highest signal when the eye movements were small, however the signal occasionally vanished in the moments of significant axial eye movements altering the optical phase. Occasional lateral eye movements were also responsible for additional artifacts in the images, in particular the images from the anterior corneal region [41]. Frames containing artefacts could be automatically removed in postprocessing.

**Comparison with clinical instrument**

Clinical OCT images from the elderly subject (for comparison with the TD-FF-OCT images) were obtained using an RTVue XR 100 (Optovue, USA). The endothelium of the same subject was photographed with a clinical specular microscope (SP-3000P, Topcon, Japan).

## Data availability

The raw data that support the findings of this study are available from the first author and the corresponding author upon request.

## Code availability

Code for the fast ocular surface segmentation and tracking is provided: https://github.com/vmazlin/Corneal-axial-tracking. Other codes that support the findings of this study are available upon reasonable request from the first author but third-party restrictions may apply (e.g. proprietary code of Lumedica, Inc).

## Acknowledgements

We are grateful to Wajdene Ghouali and Cristina Georgeon for assistance with acquiring clinical OCT and specular microscopy images. We thank Otman Sandali, Wajdene Ghouali and Benedicte Dupas for the support of the patient screening. We also thank Sylvia Desissaire for fruitful comments and suggestions on the eye tracking.

This work was supported by the HELMHOLTZ synergy grant funded by the European Research Council (ERC) (610110), a CNRS pre-maturation grant, PSL pre-maturation grant under the French Government program "Investissements d'Avenir" (ANR-10-IDEX-0001-02 PSL) and proof of concept (POC) grant funded by the European Research Council (ERC) (957546).

## Author contributions

V.M. and A.C.B. conceptualized the general idea. V.M. conceived and developed the instrument (optics, compact design, software). V.M. and K.I. performed the in vivo imaging experiments. K.G. and C.B. performed the ex vivo imaging tests. V.M., K.I. and K.G. analyzed the acquired data. V.M. and K.G. analyzed the safety and obtained the bioethics approval. K.G., V.B., C.B. and M.P. supervised the clinical organization of work at Quinze-Vingts National Eye Hospital. V.M., M.F. and A.C.B. supervised the technical side of the project. V.M. wrote the manuscript, and all the authors contributed with revisions.

# Competing interests

VM, MP, MF, KG, ACB are listed as inventors on provisional patent application that concerns the technology presented in the manuscript. Other authors declare no competing interests.

# References

1. R. R. A. Bourne, G. A. Stevens, R. A. White, J. L. Smith, S. R. Flaxman, H. Price, J. B. Jonas, J. Keeffe, J. Leasher, K. Naidoo, K. Pesudovs, S. Resnikoff, and H. R. Taylor, "Causes of vision loss worldwide, 1990–2010: a systematic analysis," Lancet Glob. Health **1**, e339–e349 (2013).
2. M. Yanoff and J. S. Duker, eds., *Ophthalmology*, Fifth edition (Elsevier, 2019).
3. M. E. Pease, S. J. McKinnon, H. A. Quigley, L. A. Kerrigan–Baumrind, and D. J. Zack, "Obstructed Axonal Transport of BDNF and Its Receptor TrkB in Experimental Glaucoma," Invest. Ophthalmol. Vis. Sci. **41**, 764–774 (2000).
4. A. Sommer, "Glaucoma: Facts and fancies," Eye **10**, 295–301 (1996).
5. C. B. Toris, M. E. Yablonski, Y.-L. Wang, and C. B. Camras, "Aqueous humor dynamics in the aging human eye," Am. J. Ophthalmol. **127**, 407–412 (1999).
6. W. M. Grant, "Further Studies on Facility of Flow Through the Trabecular Meshwork," Arch. Ophthalmol. **60**, 523–533 (1958).
7. H. A. Quigley, "The number of people with glaucoma worldwide in 2010 and 2020," Br. J. Ophthalmol. **90**, 262–267 (2006).
8. L. Mastropasqua, P. Carpineto, M. Ciancaglini, and P. E. Gallenga, "A 12-month, randomized, double-masked study comparing latanoprost with timolol in pigmentary glaucoma," Ophthalmology **106**, 550–555 (1999).
9. T. M. Richardson, "The Outflow Tract in Pigmentary Glaucoma: A Light and Electron Microscopic Study," Arch. Ophthalmol. **95**, 1015 (1977).
10. B. J. King, S. A. Burns, K. A. Sapoznik, T. Luo, and T. J. Gast, "High-Resolution, Adaptive Optics Imaging of the Human Trabecular Meshwork In Vivo," Transl. Vis. Sci. Technol. **8**, 5 (2019).
11. D. Huang, E. A. Swanson, C. P. Lin, J. S. Schuman, W. G. Stinson, W. Chang, M. R. Hee, T. Flotte, K. Gregory, C. A. Puliafito, and J. G. Fujimoto, "Optical Coherence Tomography," Science **254**, 1178–1181 (1991).
12. H. Hoerauf, "Transscleral Optical Coherence Tomography: A New Imaging Method for the Anterior Segment of the Eye," Arch. Ophthalmol. **120**, 816 (2002).
13. M. Gora, K. Karnowski, M. Szkulmowski, B. J. Kaluzny, R. Huber, A. Kowalczyk, and M. Wojtkowski, "Ultra high-speed swept source OCT imaging of the anterior segment of human eye at 200 kHz with adjustable imaging range," Opt. Express **17**, 14880 (2009).
14. B. Potsaid, B. Baumann, D. Huang, S. Barry, A. E. Cable, J. S. Schuman, J. S. Duker, and J. G. Fujimoto, "Ultrahigh speed 1050nm swept source / Fourier domain OCT retinal and anterior segment imaging at 100,000 to 400,000 axial scans per second," Opt. Express **18**, 20029 (2010).
15. I. Grulkowski, J. J. Liu, B. Potsaid, V. Jayaraman, C. D. Lu, J. Jiang, A. E. Cable, J. S. Duker, and J. G. Fujimoto, "Retinal, anterior segment and full eye imaging using ultrahigh speed swept source OCT with vertical-cavity surface emitting lasers," Biomed. Opt. Express **3**, 2733 (2012).
16. P. Li, L. An, G. Lan, M. Johnstone, D. Malchow, and R. K. Wang, "Extended imaging depth to 12 mm for 1050-nm spectral domain optical coherence tomography for imaging the whole anterior segment of the human eye at 120-kHz A-scan rate," J. Biomed. Opt. **18**, 016012 (2013).
17. M. Draelos, P. Ortiz, R. Qian, C. Viehland, R. McNabb, K. Hauser, A. N. Kuo, and J. A. Izatt, "Contactless optical coherence tomography of the eyes of freestanding individuals with a robotic scanner," Nat. Biomed. Eng. **5**, 726–736 (2021).
18. S. Chen, B. Potsaid, Y. Li, J. Lin, Y. Hwang, E. M. Moult, J. Zhang, D. Huang, and J. G. Fujimoto, "High speed, long range, deep penetration swept source OCT for structural and angiographic imaging of the anterior eye," Sci. Rep. **12**, 992 (2022).
19. K. Bizheva, N. Hutchings, L. Sorbara, A. A. Moayed, and T. Simpson, "In vivo volumetric imaging of the human corneo-scleral limbus with spectral domain OCT," Biomed. Opt. Express **2**, 1794 (2011).
20. L. Han, B. Tan, Z. Hosseinaee, L. K. Chen, D. Hileeto, and K. Bizheva, "Line-scanning SD-OCT for in-vivo, non-contact, volumetric, cellular resolution imaging of the human cornea and limbus," Biomed. Opt. Express **13**, 4007 (2022).
21. J. S. Schuman, "Spectral domain optical coherence tomography for glaucoma (an AOS thesis)," Trans. Am. Ophthalmol. Soc. **106**, 426–458 (2008).
22. W. Choi, M. woo Lee, H. G. Kang, H. S. Lee, H. W. Bae, C. Y. Kim, and G. J. Seong, "Comparison of the trabecular meshwork length between open and closed angle with evaluation of the scleral spur location," Sci. Rep. **9**, 6857 (2019).

23. T. A. Tun, M. Baskaran, C. Zheng, L. M. Sakata, S. A. Perera, A. S. Chan, D. S. Friedman, C. Y. Cheung, and T. Aung, "Assessment of trabecular meshwork width using swept source optical coherence tomography," Graefes Arch. Clin. Exp. Ophthalmol. **251**, 1587–1592 (2013).
24. J.-H. Park, H. W. Chung, E. G. Yoon, M. J. Ji, C. Yoo, and Y. Y. Kim, "Morphological changes in the trabecular meshwork and Schlemm's canal after treatment with topical intraocular pressure-lowering agents," Sci. Rep. **11**, 18169 (2021).
25. L. Kagemann, G. Wollstein, H. Ishikawa, R. A. Bilonick, P. M. Brennen, L. S. Folio, M. L. Gabriele, and J. S. Schuman, "Identification and Assessment of Schlemm's Canal by Spectral-Domain Optical Coherence Tomography," Investig. Opthalmology Vis. Sci. **51**, 4054 (2010).
26. Z. Chen, J. Sun, M. Li, S. Liu, L. Chen, S. Jing, Z. Cai, Y. Xiang, Y. Song, H. Zhang, and J. Wang, "Effect of age on the morphologies of the human Schlemm's canal and trabecular meshwork measured with swept-source optical coherence tomography," Eye **32**, 1621–1628 (2018).
27. P. Li, A. Butt, J. L. Chien, M. P. Ghassibi, R. L. Furlanetto, C. F. Netto, Y. Liu, W. Kirkland, J. M. Liebmann, R. Ritch, and S. C. Park, "Characteristics and variations of in vivo Schlemm's canal and collector channel microstructures in enhanced-depth imaging optical coherence tomography," Br. J. Ophthalmol. **101**, 808–813 (2017).
28. E. Beaurepaire, A. C. Boccara, M. Lebec, L. Blanchot, and H. Saint-Jalmes, "Full-field optical coherence microscopy," Opt. Lett. **23**, 244–246 (1998).
29. V. Mazlin, P. Xiao, E. Dalimier, K. Grieve, K. Irsch, J.-A. Sahel, M. Fink, and A. C. Boccara, "In vivo high resolution human corneal imaging using full-field optical coherence tomography," Biomed. Opt. Express **9**, 557 (2018).
30. M. Flocks, "The Anatomy of the Trabecular Meshwork as Seen in Tangential Section," Arch. Ophthalmol. **56**, 708–718 (1956).
31. V. Mazlin, P. Xiao, J. Scholler, K. Irsch, K. Grieve, M. Fink, and A. C. Boccara, "Real-time non-contact cellular imaging and angiography of human cornea and limbus with common-path full-field/SD OCT," Nat. Commun. **11**, 1868 (2020).
32. V. Mazlin, K. Irsch, M. Paques, J.-A. Sahel, M. Fink, and C. A. Boccara, "Curved-field optical coherence tomography: large-field imaging of human corneal cells and nerves," Optica **7**, 872 (2020).
33. P. Mecê, K. Groux, J. Scholler, O. Thouvenin, M. Fink, K. Grieve, and C. Boccara, "Coherence gate shaping for wide field high-resolution in vivo retinal imaging with full-field OCT," Biomed. Opt. Express **11**, 4928 (2020).
34. O. Thouvenin, M. Fink, and A. C. Boccara, "Dynamic multimodal full-field optical coherence tomography and fluorescence structured illumination microscopy," J. Biomed. Opt. **22**, 1 (2017).
35. P. Xiao, V. Mazlin, K. Grieve, J.-A. Sahel, M. Fink, and A. C. Boccara, "In vivo high-resolution human retinal imaging with wavefront-correctionless full-field OCT," Optica **5**, 409 (2018).
36. A. Dubois, *Handbook of Full-Field Optical Coherence Microscopy: Technology and Applications* (Pan Stanford publishing, 2016).
37. K. Grieve, D. Ghoubay, C. Georgeon, G. Latour, A. Nahas, K. Plamann, C. Crotti, R. Bocheux, M. Borderie, T.-M. Nguyen, F. Andreiuolo, M.-C. Schanne-Klein, and V. Borderie, "Stromal striae: a new insight into corneal physiology and mechanics," Sci. Rep. **7**, (2017).
38. M. J. Hogan, J. A. Alvarado, and J. E. Weddell, *Histology of the Human Eye: An Atlas and Textbook* (Saunders, 1971).
39. J. M. Gonzalez, M. K. Ko, A. Pouw, and J. C. H. Tan, "Tissue-based multiphoton analysis of actomyosin and structural responses in human trabecular meshwork," Sci. Rep. **6**, 21315 (2016).
40. O. Masihzadeh, D. A. Ammar, M. Y. Kahook, E. A. Gibson, and T. C. Lei, "Direct trabecular meshwork imaging in porcine eyes through multiphoton gonioscopy," J. Biomed. Opt. **18**, 036009 (2013).
41. V. Mazlin, P. Xiao, K. Irsch, J. Scholler, K. Groux, K. Grieve, M. Fink, and A. C. Boccara, "Optical phase modulation by natural eye movements: application to time-domain FF-OCT image retrieval," Biomed. Opt. Express **13**, 902 (2022).
42. W. H. Spencer, J. Alvarado, and T. L. Hayes, "Scanning electron microscopy of human ocular tissues: trabecular meshwork," Invest. Ophthalmol. **7**, 651–662 (1968).
43. J. W. Rohen, R. Futa, and E. Lütjen-Drecoll, "The fine structure of the cribriform meshwork in normal and glaucomatous eyes as seen in tangential sections," Invest. Ophthalmol. Vis. Sci. **21**, 574–585 (1981).
44. C. M. Marando, C. Y. Park, J. A. Liao, J. K. Lee, and R. S. Chuck, "Revisiting the Cornea and Trabecular Meshwork Junction With 2-Photon Excitation Fluorescence Microscopy," Cornea **36**, 704–711 (2017).
45. C. Y. Park, J. K. Lee, M. Y. Kahook, J. S. Schultz, C. Zhang, and R. S. Chuck, "Revisiting Ciliary Muscle Tendons and Their Connections With the Trabecular Meshwork by Two Photon Excitation Microscopic Imaging," Investig. Opthalmology Vis. Sci. **57**, 1096 (2016).
46. "Glaucoma - UK National Screening Committee (UK NSC) - GOV.UK," https://view-health-screening-recommendations.service.gov.uk/glaucoma/.
47. V. A. Moyer, "Screening for Glaucoma: U.S. Preventive Services Task Force Recommendation Statement," Ann. Intern. Med. (2013).
48. M. Pircher and R. J. Zawadzki, "Review of adaptive optics OCT (AO-OCT): principles and applications for retinal imaging [Invited]," Biomed. Opt. Express **8**, 2536 (2017).
49. R. S. Jonnal, O. P. Kocaoglu, R. J. Zawadzki, Z. Liu, D. T. Miller, and J. S. Werner, "A Review of Adaptive Optics Optical Coherence Tomography: Technical Advances, Scientific Applications, and the Future," Investig. Opthalmology Vis. Sci. **57**, OCT51 (2016).

50. Z. Liu, K. Kurokawa, F. Zhang, J. J. Lee, and D. T. Miller, "Imaging and quantifying ganglion cells and other transparent neurons in the living human retina," Proc. Natl. Acad. Sci. **114**, 12803–12808 (2017).
51. S. Desissaire, A. Pollreisz, A. Sedova, D. Hajdu, F. Datlinger, S. Steiner, C. Vass, F. Schwarzhans, G. Fischer, M. Pircher, U. Schmidt-Erfurth, and C. K. Hitzenberger, "Analysis of retinal nerve fiber layer birefringence in patients with glaucoma and diabetic retinopathy by polarization sensitive OCT," Biomed. Opt. Express **11**, 5488 (2020).
52. M. Rispoli, E. Souied, D. Huang, and B. Lumbroso, *Practical Handbook of Oct Angiography* (Jaypee, 2016).
53. L. Puyo, M. Paques, and M. Atlan, "Retinal blood flow reversal quantitatively monitored in out-of-plane vessels with laser Doppler holography," Sci. Rep. **11**, 17828 (2021).
54. L. Puyo, C. David, R. Saad, S. Saad, J. Gautier, J. A. Sahel, V. Borderie, M. Paques, and M. Atlan, "Laser Doppler holography of the anterior segment for blood flow imaging, eye tracking, and transparency assessment," Biomed. Opt. Express **12**, 4478 (2021).
55. C. Pfäffle, H. Spahr, L. Kutzner, S. Burhan, F. Hilge, Y. Miura, G. Hüttmann, and D. Hillmann, "Simultaneous functional imaging of neuronal and photoreceptor layers in living human retina," Opt. Lett. **44**, 5671 (2019).
56. J. Scholler, K. Groux, K. Grieve, C. Boccara, and P. Mecê, "Adaptive-glasses time-domain FFOCT for wide-field high-resolution retinal imaging with increased SNR," Opt. Lett. **45**, 5901 (2020).
57. O.-Y. Tektas and E. Lütjen-Drecoll, "Structural changes of the trabecular meshwork in different kinds of glaucoma," Exp. Eye Res. **88**, 769–775 (2009).
58. B. Liu, S. McNally, J. I. Kilpatrick, S. P. Jarvis, and C. J. O'Brien, "Aging and ocular tissue stiffness in glaucoma," Surv. Ophthalmol. **63**, 56–74 (2018).
59. J. Buffault, A. Labbé, P. Hamard, F. Brignole-Baudouin, and C. Baudouin, "The trabecular meshwork: Structure, function and clinical implications. A review of the literature," J. Fr. Ophtalmol. **43**, e217–e230 (2020).
60. S. Chen, X. Liu, N. Wang, X. Wang, Q. Xiong, E. Bo, X. Yu, S. Chen, and L. Liu, "Visualizing Micro-anatomical Structures of the Posterior Cornea with Micro-optical Coherence Tomography," Sci. Rep. **7**, 10752 (2017).
61. A. Wartak, M. S. Schenk, V. Bühler, S. A. Kassumeh, R. Birngruber, and G. J. Tearney, "Micro-optical coherence tomography for high-resolution morphologic imaging of cellular and nerval corneal micro-structures," Biomed. Opt. Express **11**, 5920 (2020).
62. B. Tan, Z. Hosseinaee, L. Han, O. Kralj, L. Sorbara, and K. Bizheva, "250 kHz, 1,5 μm resolution SD-OCT for in-vivo cellular imaging of the human cornea," Biomed. Opt. Express **9**, 6569 (2018).
63. E. Auksorius, D. Borycki, P. Stremplewski, K. Liżewski, S. Tomczewski, P. Niedźwiedziuk, B. L. Sikorski, and M. Wojtkowski, "In vivo imaging of the human cornea with high-speed and high-resolution Fourier-domain full-field optical coherence tomography," Biomed. Opt. Express **11**, 2849 (2020).
64. X. Yao, K. Devarajan, R. M. Werkmeister, V. A. dos Santos, M. Ang, A. Kuo, D. W. K. Wong, J. Chua, B. Tan, V. A. Barathi, and L. Schmetterer, "In vivo corneal endothelium imaging using ultrahigh resolution OCT," Biomed. Opt. Express **10**, 5675 (2019).
65. C. Canavesi, A. Cogliati, A. Mietus, Y. Qi, J. Schallek, J. P. Rolland, and H. B. Hindman, "In vivo imaging of corneal nerves and cellular structures in mice with Gabor-domain optical coherence microscopy," Biomed. Opt. Express **11**, 711 (2020).
66. T. D. Weber and J. Mertz, "In vivo corneal and lenticular microscopy with asymmetric fundus retroillumination," Biomed. Opt. Express **11**, 3263 (2020).
67. R. F. Guthoff, A. Zhivov, and O. Stachs, "*In vivo* confocal microscopy, an inner vision of the cornea - a major review," Clin. Experiment. Ophthalmol. **37**, 100–117 (2009).
68. C. Xin, M. Johnstone, N. Wang, and R. K. Wang, "OCT Study of Mechanical Properties Associated with Trabecular Meshwork and Collector Channel Motion in Human Eyes," PLOS ONE **11**, e0162048 (2016).
69. G. Li, C. Lee, V. Agrahari, K. Wang, I. Navarro, J. M. Sherwood, K. Crews, S. Farsiu, P. Gonzalez, C.-W. Lin, A. K. Mitra, C. R. Ethier, and W. D. Stamer, "In vivo measurement of trabecular meshwork stiffness in a corticosteroid-induced ocular hypertensive mouse model," Proc. Natl. Acad. Sci. **116**, 1714–1722 (2019).
70. C. Xin, S. Song, M. Johnstone, N. Wang, and R. K. Wang, "Quantification of Pulse-Dependent Trabecular Meshwork Motion in Normal Humans Using Phase-Sensitive OCT," Investig. Opthalmology Vis. Sci. **59**, 3675 (2018).
71. S. Kim, M. Crose, W. J. Eldridge, B. Cox, W. J. Brown, and A. Wax, "Design and implementation of a low-cost, portable OCT system," Biomed. Opt. Express **9**, 1232 (2018).
72. S. Patel and L. Tutchenko, "The refractive index of the human cornea: A review," Contact Lens Anterior Eye **42**, 575–580 (2019).
73. S. Labiau, G. David, S. Gigan, and A. C. Boccara, "Defocus test and defocus correction in full-field optical coherence tomography," Opt. Lett. **34**, 1576 (2009).